# Observation of protected localized states induced by curved space in acoustic topological insulators

Hong-Wei Wu,[1,2,*] Jia-Qi Quan,[1] Yun-Kai Liu,[1] Yiming Pan,[3] Zong-Qiang Sheng[1] & Lai-Wang Jing[1]

[1]*School of Mechanics and Photoelectric Physics, Anhui University of Science and Technology, Huainan 232001, China*
[2]*National Laboratory of Solid State Microstructures, Nanjing University, Nanjing 210093, China*
[3]*Physics Department and Solid State Institute, Technion, Haifa 32000, Israel*

Topological insulators (TIs) with robust boundary states against perturbations and disorders provide a unique approach for manipulating waves, whereas curved space can effectively control the wave propagation on curved surfaces by the geometric potential effect as well. In general, two-dimensional (2D) TIs are designed on a flat surface; however, in most practical cases, curved topological structures are required. In this study, we design a 2D curved acoustic TI by perforation on a curved rigid plate. We experimentally demonstrate that a topological localized state stands erect in the bulk gap, and the corresponding pressure distributions are confined at the position with the maximal curvature. Moreover, we experimentally verify the robustness of the topological localized state by introducing defects near the localized position. To understand the underlying mechanism of the topological localized state, a tight-binding model considering the geometric potential effect is proposed. The interaction between the geometrical curvature and topology in the system provides a novel scheme for manipulating and trapping wave propagation along the boundary of curved TIs, thereby offering potential applications in flexible devices.

*Corresponding author: hwwu@aust.edu.cn

Topology, which is a mathematical concept for describing the space-preserved property in continuous deformation, has been introduced into condensed-matter physics to investigate various phenomena, including the quantum integer Hall effect[1,2], quantum spin Hall effect[3,4], and topological insulators (TIs)[5,6]. In the past several decades, TIs have constituted an expanding research field in condensed matter, and the robust transport effect of boundary states against disorders has attracted intense interest in classical systems[7-12]. Most recently, with the intensive investigations into TIs, higher-order TIs (HOTIs)[13,14] with unconventional bulk–boundary correspondence has been proposed to support the lower-dimensional topological boundary states, thereby opening a new research direction in classical waves[15-29]. The rule can be described as a $d$-dimensional ($d$D) TI with ($d$-1)D, …, ($d$-$n$-1)D gapped boundary states and ($d$-$n$)D gapless boundary states for an $n$th-order TI. Owing to the flexibility of the sample design in classical wave systems, 2D HOTIs have been implemented extensively in photonics[17-20], electrical circuits[21,22], and acoustic systems[23-29]. Topological protected edge and corner states in 2D HOTIs can provide an avenue for designing novel functional components to increase the integration density and reduce the noise disturbance resulting from fabrication imperfections on integrated chips.

As another interesting topic in wave propagation, curved space has been studied extensively for defining the transported properties of quantum particles[30-32] and classical waves[33-45] that are confined on a curved surface. As early as 1981, pioneering work presented by da Costa[32] indicated that the motion of a quantum particle constrained on a curved surface by a physical confining potential experiences effective geometric potential. With reference to the similarity between the Schrödinger equation and paraxial Helmholtz equation, optical analogues of the quantum geometric potential have been studied theoretically and experimentally in optical systems for light wave packets constrained on thin dielectric guiding layers[34-40]. With the aid of the geometric potential effect, numerous quantum phenomena have been migrated to classical systems, such as optical Bloch oscillation[33,34], Zener tunneling[37,38], and the Wolf effect[40]. In recent years, a curved surface plasmon polariton (SPP) system was proposed to investigate the geometric potential effect[42-44]. Unlike photons that are squeezed on a curved surface by a thin dielectric layer, SPPs are intrinsically 2D quasi-particles without any squeezed potential requirements. Based on this versatile platform, the geometric potential effect has been experimentally demonstrated to manipulate SPP propagation along a curved interface between metal and dielectric materials.

Furthermore, with the research upsurge in TIs, geometric potentials have been used to explore topological phases in curved-space systems. For example, by setting a cylindrical thin waveguide layer with an imprinted curved Su–Schrieffer–Heeger (SSH) binary lattice, Segev et. al[45] demonstrated that the space curvature can induce topological edge states, topological phase transitions, Thouless pumping, and localization effects. In particular, recent noteworthy work revealed photonic topological phases of matter in the hyperbolic geometry of non-Euclidean space[46]. The topological hyperbolic lattice is designed by mapping Euclidean Bravais lattices to a hyperbolic surface in non-Euclidean space, following which the topological protection of helical edge states is demonstrated by measuring the edge confinement and defect immunity. In summary, opportunities exist for investigating the topological properties of crystals in non-Euclidean space, and it is desirable to gain new insights into TIs.

In this work, we designed a curved 2D second-order TI in an acoustic system by perforating the curved surface of a rigid wall based on the lattice distribution of the 2D SSH model. We experimentally observed that a topological edge state with the highest eigenfrequency is distinct from the edge states and exhibits strength mode localization induced by the geometric curvature. By introducing near and far defects into the structure, we also demonstrated that the topological localized state is robust against disturbances. Furthermore, the tight-binding model considering the geometric potential effect was used to reveal the physical mechanism underlying the topological localized state. The results indicate that the nonhomogeneous onsite energy leads to mode localization. Although our study was carried out in the context of acoustics, the concepts involved are universal, with manifestations in many branches of physics, including electronic, electromagnetic and elastic systems. Curved TIs provide the unique ability to manipulate and trap the waves propagated along the edges of TIs, and offer potential applications in flexible devices.

## Results

**Construction of curved 2D second-order TI.** Our 2D curved second-order TI was designed by perforating a curved surface of a rigid wall to form a square lattice distribution, as illustrated in Fig. 1a. The yellow region represents the curved rigid wall with a length of 65.8 cm, width of 48.4 cm, and thickness of 3.5 cm. The shape of the bump could be described by the function $z = H \cos(\pi x/L)$, where the bump height $H = 5$ cm, bump range $-L/2 < x < L/2$, and $L = 40$ cm. The depth and radius of the holes were $h = 3$ cm and $r = 0.3$ cm, respectively, as indicated in the right inset. The lattice constant was $a = 4$ cm. The distance between the neighboring holes in a unit cell was represented as $d$, and the topological phases could be changed by tuning the ratio $d/a$. It should be explained why the 2D curved second-order TI was designed using the perforated method. The reason is that the holey structure could support the spoof acoustic surface waves confined on the curved surface to satisfy the condition of generating the geometric potential effect. Such holey structures are common in acoustic metamaterials. Previous work based on this type of holey structure on a 2D flat surface realized acoustic topological corner states[26]. In our work, the entire rigid wall included two types of hole arrays ($d/a = 0.75$ and $0.25$) to form an interface, as indicated by the bright trace in Fig. 1a. The two insets on the right of Fig. 1a depict the air domain of the unit cell in the cyan region.

Prior to investigating the topological properties of 2D curved second-order TIs, we recall the band structure of the square lattice distributed on a flat surface corresponding to $H = 0$ cm, and demonstrate the band inversion and topological phase transition when changing the ratio $d/a$ in Fig. 1b. The red pentagrams show the band diagram of the square lattice for the ratio $d/a = 0.5$. The first Brillouin zone is displayed in the inset of Fig. 1b. The blue dashed lines represent the dispersion relationships for the sound propagating in the air background. When comparing the band structure and air line, the feature of spoof acoustic surface waves could be observed[26]. Furthermore, the bands $s$ and $p_x$ were degenerated along the X–M direction for the ratio $d/a = 0.5$. When the distance between the four holes in the unit cell was decreased to $d/a = 0.25$ or increased to $d/a = 0.75$, the degenerate band was separated to form a bandgap, as indicated by the black dots. To demonstrate the topological phase transition from trivial to nontrivial by increasing the distance from $d/a = 0.25$ to $0.75$, we first present the parities of the pressure fields in the unit cell at point X, as illustrated in Fig. 1c, for calculating the 2D Zak phases. It is obvious that the eigenstate with a high frequency possessed odd parity corresponding to the $p_x$ band, whereas the eigenstate with a low frequency had even parity corresponding to the $s$ band for $d/a = 0.25$. When increasing the distance between the four holes in the unit cell to $d/a = 0.75$, the energy bands were inversed, as can be observed in Fig. 1c. To characterize the topological phase transition rigorously, the 2D Zak phases were calculated by the parities at high symmetric points, as follows[26]:

$$P_m = \frac{1}{2}\left(\sum_n q_m^n \bmod 2\right), \quad (-1)^{q_m^n} = \frac{\eta(X_m)}{\eta(\Gamma)}, \tag{1}$$

where $\eta$ denotes the parity associated with $\pi$ rotation at high symmetric points and $m$ represents the $x$ or $y$ direction. The summation $\sum_n q_m^n$ was taken over all bands from the 1st to $n$th. For the square lattice with $C_{4v}$ symmetry, the Zak phase $P_x$ was identical to $P_y$ in the y direction; that is, $P_x = P_y$. According to Eq. (1) and all eigenstates at high symmetric points, we could obtain the Zak phases $\boldsymbol{P} = (0, 0)$ for the ratio $d/a < 0.5$ and $\boldsymbol{P} = (1/2, 1/2)$ for the ratio $d/a > 0.5$. The results reveal that the structures with ratios of $d/a < 0.5$ and $d/a > 0.5$ corresponded to topological nontrivial and trivial insulating phases, respectively. We demonstrated the generation of nontrivial dipole moments in both the $x$ and $y$ directions of the second-order TIs. When the two dipole moments $P_x$ and $P_y$ converged and form a quadruple tensor at the corner, a corner state defined by a topological corner charge[26] $Q_c = 4P_xP_y$ could be induced and protected by the edge–corner correspondence. For example, the charge was $Q_c = 1$ at the corner of the topological nontrivial structure.

Subsequently, we investigated the influence of the geometrical curvature on the band structure of the 2D curved second-order TI for the square lattices distributed on a curved surface. As a reference, we constructed a square crystal composed of a $6a \times 16a$ nontrivial region with $d/a = 0.75$ near a same-sized trivial region with $d/a = 0.25$ on the flat holey plate to form an interface, as illustrated in Fig. 2a. Figure 2b presents the curved case, which had the same lattice distribution as the flat holey plate, with 480 holes located on the bump surface. Figures 2c and 2d depict the discrete eigenfrequencies for the flat and curved structures, respectively. In this simulation, we set the boundary conditions around the entire structure as plane wave radiation boundaries to focus only on the edge states along the interface between the topological trivial and nontrivial parts for a clear observation of the influence of the geometrical curvature on the edge states. The difference between the radiation boundary condition and air area around the structure is illustrated in Fig. S1 of the Supplementary Material. It can be observed that there are 15 edge states supported by the interface presented in Fig. 2c, which are marked by red dots. These edge states are clarified by their eigenstates in Fig. S2 of the Supplementary Material. For the curved case, as illustrated in Fig. 2d, the number of edge states was not changed, but the eigenfrequencies of the edge states were obviously increased. By determining the mode distributions of these eigenstates corresponding to the eigenfrequencies, as indicated in Fig. S2, an interesting phenomenon was observed whereby a localized state emerged at the top of the bump, as shown in the inset of Fig. 2d. The eigenfrequency of this localized state is marked by a blue dot in Fig. 2d. The mode distributions of the topological edge states corresponding to the 12th edge states are depicted in the insets of Figs. 2c and 2d. Comparing the eigenfrequency distributions in Fig. 2c and 2d, the edge states exhibited energy degeneration in the curved case, which is verified by the mode distributions in c2 to l2 of Fig. S2 in the Supplementary Material. The reason is that the bilaterally symmetrical geometrical shape led to bilaterally symmetrical distributions of the onsite energy of the holes on the curved surface; consequently, the eigenstates of the holey structure located at the two slopes underwent energy degeneration. In particular, the eigenstate located at the top of the bump with the maximal onsite energy was isolated from the other degenerated edge states to form a topological localized state. Furthermore, the evolution of the eigenfrequency when increasing the height $H$ of the curved structure is illustrated in Fig. S3 in the Supplementary Material. A tight-binding model is presented in the following theoretical section to clarify the physical mechanism.

**Observation of topological localized state.** Next, we experimentally measured the topological localized state in a curved second-order TI. The holey structure was constructed via 3D printing of a curved epoxy resin plate and perforation on the curved plate surface. Epoxy resin can be treated as a rigid wall compared to air. Figure 3a presents the measurement setup. The sample was surrounded by absorbing sponges and a point source was placed at a distance of 3 cm from the curved surface of the sample. A microphone was placed at the back of the sample to detect the pressure amplitudes. Figure 3b indicates the placements of the source and microphone, marked by colored pentagrams and circles, respectively, for measuring the pressure amplitudes of the bulk, edge, and localized states. The transmission spectra of these eigenstates are depicted in Fig. 3c. It can be observed that the transmission peaks of the edge and localized states were clearly located in the bandgap of the bulk band. In particular, the frequency $f =$ 2661 Hz of the topological localized state agreed very well with the simulation frequency $f = 2660.7$ Hz. It should be highlighted that the localized state was topologically protected because it was located in the bulk gap of the topological nontrivial structure. To observe the topological localized state in real space clearly, we also measured the pressure distributions of all holes on the curved surface at $f$ = 2661 Hz in Fig. 3d. The result demonstrated the field confinement of the topological localized state at the structural top with the maximal curvature. Furthermore, to compare the measured and simulated results, we scanned the pressured field amplitude profile in each hole along two directions, as indicated by the cyan and orange dashed lines marked with b and c in Fig. 4a. Figures 4b and 4c present the normalized pressure field distributions along the b and c directions, where the black solid lines represent the distribution in the simulation and the red dotted dashed lines correspond to the measured results. The measured results indicate that the pressure fields were mainly localized at the holes of the top, which were consistent with the simulated results.

To demonstrate the robustness of the topological localized state against defects, we further introduced near and far defects into the curved structure by inserting solid cylinders into the holes, as indicated by the cyan and orange dashed circles in Fig. 4d. The bottom-right inset clearly shows the positions of the two types of defects. Figure 4e displays the transmission spectra of the topological localized state with and without the near defects. The black solid line is the transmission spectrum of the perfectly curved structure and the transmission peak was located at $f$ = 2661 Hz. When introducing the near defects into the curved structure, the magnitude of the peak was impregnable and the topological localized state remained within the bandgap with negligible red-shifting, as indicated by the blue dashed line in the simulation and red dashed line in the experiment. Figure 4f presents a similar phenomenon for the introduction of the far defects. These results suggest that the topological localized state was robust against near and far defects. In fact, the robustness of the topological localized state was natural because the root of this state was a topological localized state protected by bulk–edge correspondence.

**Tight-binding model considering geometric potential effect.** To understand the physical mechanism of the topological localized state induced by the curved space, we theoretically calculated the eigenfrequencies of the edge states based on a tight-binding model by considering the geometric potential of the curved space. For this purpose, we first performed conformal transformations between the Cartesian coordinates ($x$, $y$, $z$) and curvilinear coordinates ($v$, $y$, $u$) indicated in Fig. 5a using the relational expression $z + ix = R(v) \exp[(u + iv) / R(v)]$, where $R(v)$ represents the radius of the curvature varied with the curvilinear coordinate $v$. According to the conformal transformation, and considering that the spoof acoustic surface waves only propagated along the interface between the topological trivial and nontrivial parts owing to the confinement of the bulk bandgap, the spoof acoustic surface wave was expressed as $p(v, u) = A(v, u)\exp(i\beta v)$ under paraxial approximation, where $A(v, u)$ is the slowly varying amplitude and $\beta$ is the wavevector. By applying the slowly varying envelope approximation to the 2D scalar wave equation, the acoustic surface wave yielded the acoustic Schrödinger equation[37,44]:

$$i\hbar \frac{\partial A}{\partial v} = -\frac{\hbar^2}{2n_{sw}}\frac{\partial^2 A}{\partial u^2} + \frac{n_{sw}^2 - n^2(v,u)}{2n_{sw}}A - \frac{un^2(v,u)}{R(v)n_{sw}}A, \qquad (2)$$

where $\hbar = 1/k_0$ and $k_0$ is the wavevector of free space. Further details can be found in the Supplementary Material. In this case, $n(v, u) = k(v, u)/k_0$ corresponds to the refractive index of the holey materials, and $n_{sw} = \beta/k_0$. For the spoof acoustic surface waves confined on the curved surface, the second term on the right of Eq. (2) corresponds to the confining potential $V_c (v, u) = [n_{sw}^2 - n^2(v, u)]/(2n_{sw})$. The final term is the geometric potential[37] $V_g (v, u) = - un^2(v, u)/R(v)n_{sw}$. The geometric potential is mainly dependent on the curvature $R(v)$ and can be expressed as $V_g (v) = - \alpha/R(v)$ for the spoof surface wave propagated on the curved surface, with $\alpha = un^2(v, u)/n_{sw}$ as a parameter. For simplicity, we set the value as $\alpha = 10$.

For our structure described by the function $z = H \cos(\pi x/L)$, we calculated the geometric potential for different heights $H$ along the curvilinear coordinate $v$ when the span $L = 40$ cm, as illustrated in Fig. 5b. It is obvious that the geometric potential was 0 for $H = 0$ cm, corresponding to a flat structure. When increasing the height $H$ to 2.5 cm, the geometric potential gradually increased along the left structural slope, and the maximal value was attained at the top of the bump with the maximal curvature. Thereafter, the geometric potential gradually decreased from the maximal value to 0 along the right slope. When increasing the height $H$ to 7.5 cm, a higher potential barrier was formed along the curvilinear coordinate $v$. Conversely, when changing the sign of the height $H$ to negative, a potential well was formed for the concave surface, as indicated in Fig. 5b. To clarify the influence of the curved span $L$ on the geometric potential, we also calculated the geometric potentials when decreasing the span from $L = 60$ cm to 30 cm with a constant height of $H = 5$ cm, as depicted in Fig. 5c. It can be observed that the potential barrier was increased when decreasing the span. For the concave surface, the depth of the potential well decreased when decreasing the span $L$; this result is not provided here.

It is well known that topological edge states are well confined in 1D holes at the interface between the topological nontrivial and trivial parts owing to the bulk–boundary correspondence. Thus, we performed tight-binding approximation to calculate the eigenfrequencies of the 1D hole array at the curved interface by introducing discrete geometric potentials into the onsite energy (that is, resonant frequency) of the holes. It should be noted that the resonant frequency of the hole was 2685 Hz, and the coupling coefficients between two holes were 50 Hz for the ratio $d/a$ = 0.25 and 15 Hz for $d/a$ = 0.75 obtained by the simulation. When the structure exhibited geometric curvature, the onsite energy of the holes could be modified by adding relevant discrete geometric potentials. Coupling coefficients are recognized as invariable for a weak curvature. Figure 5d depicts the eigenfrequencies of the 1D hole array for different heights of $H$ = -5 cm, 0 cm, and 5 cm when the span was $L$ = 40 cm. The black dots represent the bulk states of the 1D flat hole array. Notably, the bulk states in the 1D hole array of the theoretical model corresponded to the edge states of the 2D topological nontrivial hole array, as indicated in Fig. 5a[27]. Comparing the bulk states (black dots) in Fig. 5d and the edge states (red dots) in Fig. 2c, it can be observed that the eigenfrequencies of the bulk states in the 1D hole array were consistent with those of the edge states of the 2D hole array. The correspondence between the bulk states of the 1D hole array and edge states of the 2D hole array could also be verified by comparing their mode distributions, as illustrated in c1 to q1 of Figs. S2 and S4. In particular, the boundary states of the 1D hole array owing to the open boundary conditions are presented in the dashed box of Fig. 5d, which are marked as "BS". However, the corresponding corner states in the 2D hole array are not presented in Fig. 2c owing to the use of radiation boundary conditions.

The above discussion has demonstrated that the tight-binding model of a 1D hole array is valid for studying the edge states of a 2D hole array along the interface in a curved space. According to Fig. 5d, when the curved height was increased to $H$ = 5 cm, the eigenfrequencies of the bulk states were increased universally and the eigenstates with low frequencies resulted in energy degeneration. All of the mode distributions of the eigenstates are depicted in Fig. S5. However, the boundary states overlapped in the case of $H$ = 0 at 2685 Hz (the resonant frequency of the hole). This is because the holes supported the boundary states located at the flat regions at the two ends, the onsite energy of which was not modified by the geometric potential. As illustrated in Fig. 2d, the curved holey structure was bilaterally symmetric and the eigenstates supported on the bilateral slopes exhibited energy degeneration. Subsequently, when the bulk state (labelled with the red "LS" in Fig. 5d) with the maximal eigenfrequency was isolated, its mode was confined at the top of the curved structure with the maximal geometric potential, as can be observed in Fig. S5 of the Supplementary Material. Furthermore, the eigenfrequencies of the concave structure when the curved height was set to $H$ = -5 cm are denoted by the blue dots in Fig. 5d. It can easily be observed that the eigenfrequencies were decreased when adding a negative geometric potential (as indicated in Fig. 5b) to the resonant frequencies of the holes. Interestingly, the localized state emerged at the bottom of the curved structure with the minimum eigenfrequency, which is marked by the blue "LS" in Fig. 5d. Similarly, the eigenstates with higher frequencies exhibited energy degeneration. All of the eigenstates are depicted in Fig. S6 of the Supplementary Material. Figure 5e presents the intensity distributions of the localized state, marked by the red "LS" in the curved structure ($H$ = 5 cm), and the reference state, marked by "R1" in the flat structure. The pressure fields of the topological localized state were effectively confined at the structural top compared to the field distribution of the reference state "R1". We also compared the intensity distributions of the localized state, marked by the blue "LS" in the concave structure ($H$ = -5 cm), and the reference state, marked by "R2" in the flat structure, as illustrated in Fig. 5f.

    The proposed theoretical model suggested that the positions of the holes on the curved surface affected the distribution of the eigenstates, and the curved directions (convex and concave) determined the eigenfrequency of the topological localized state. To verify the theoretical predictions, we also simulated the curved structures with different curved directions and different hole distributions, as depicted in Fig. 6. The structures in Figs. 6a and 6b had the same lattice distribution but different curved directions, whereas the structures in Figs. 6a and 6c had the same curved direction but different lattice distributions. In Figs. 6a to d, 16$a$ and 15$a$ denote 16 cells and 15 cells along the curved surface, respectively. The top right insets display the details of the structural top. Figures 6e to h present the eigenfrequencies of the 2D curved TI structures. The topological localized state had the maximum frequency in the bandgap for the convex structure, but the minimum frequency in the bandgap for the concave structure. In general, a convex structure expands the highest edge state to form a

topological localized state, whereas a concave structure reduces the lowest edge state to form a topological localized state. The results were consistent with our theoretical predictions. Furthermore, comparing the eigenfrequencies of the curved structures in Figs. 6e and 6g (or Figs. 6f and 6h), it can be observed that the eigenfrequencies exhibited a small frequency shift owing to the location shift of the holes on the curved surface in the same curved direction. Figures 6i to l depict the intensity distributions of the topological localized states for the different cases.

**Discussion**

We developed a 2D curved second-order TI by perforating a curved rigid surface, and demonstrated that the topological localized state can stand out from the topological edge states with an increasing geometrical curvature. Our experiments directly revealed the topological localized state at the top of the curved structure, and agreed very well with the simulation. Moreover, we demonstrated that the topological localized state is robust against defect disturbance. To understand the physical mechanism, we theoretically derived the geometric potential distribution on the curved surface based on conformal transformation, and implemented the tight-binding model with modified onsite energy by considering the discrete geometric potentials. The theoretical results explained the origin of the topological localized state, in which the geometric potentials modulated the onsite energy of the holes on the curved surface to form a bilaterally symmetric distribution. Subsequently, the edge state with the highest frequency was isolated from the other degenerate edge states. Furthermore, we theoretically demonstrated that negative geometric potentials of the concave structure would induce the topological localized state at the bottom of the curved structure, which originated from the isolated edge state with the minimum eigenfrequency. These results were verified by the numerical simulation.

The mechanism of the topological localized state in a 2D curved second-order TI induced by geometrical curvature is very different from the topological corner state. The mode intensity, spatial position, and eigenfrequency of the topological localized state can be controlled by tuning the geometrical shape of the curved structure. In this work, we have investigated the topological localized state induced by only one convex or concave surface. The curved structure can be expanded to include multiple convex, multiple concave, or mixed surfaces to realize numerous topological localized states. Furthermore, it should be emphasized that our study was carried out within the context of acoustics, but the concept of the topological localized state can be generalized to various branches of physics, including electromagnetic and elastic systems. In summary, we have proposed a novel approach to manipulate the topological edge states from the propagated mode to the localized mode controlled by the geometric curvature. Our results have potential practical applications in flexible devices.

**Method**

**Simulations**. Numerical simulations in this work are all performed using the 3D acoustic module of a commercial finite-element simulation software (COMSOL Multiphysics). The resin blocks are treated as acoustically rigid materials. The mass density and sound velocity in air are taken as 1.21 kg/m$^3$ and 346m/s, respectively. In the eigen evaluations, four boundaries of the unit cells are set as Floquet periodic boundaries for the data in Fig. 1b and 1c. To suppress the edge states and corner states along the outside boundaries of the structure, the boundaries of the whole structure are set as plane wave radiation boundaries except the rigid bottom, for the data in Figs. 2, 6 and S3.

**Experiments**. In order to verify ours design, the proposed structure was precision-fabricated using epoxy resin via 3D printing. The fabricated sample in Fig.3a consisted of 768 holes with the depth 3cm and radius 0.3cm. The sample was set up in the free space which is surrounded by the absorbing sponges. A point-like sound source was designed by connecting one loudspeaker to the metallic pipes, which was placed near the structure of the holes at a distance of 3cm. The structural local pressure fields were measured by inserting the microphone into the bottom of hole. The bottom will be blocked after measure to keep the existence of spoof acoustic surface wave supported in the structure. Then the digitizer acquired the outputs of the microphones. The transmission spectra of the bulk, edge, and localized state were obtained by measuring transmission pressure fields. The indoor temperature was strictly controlled within 25±0.3°C during the whole experiment. Ambient noise is lower than 20dB, the influence on the experiment can be negligible.

**Data availability**

The data that support the plots within this work and other related finding are available from the corresponding authors upon reasonable request.

## Acknowledgements

This work was supported by the National Natural Science Foundation of China (NSFC) (Grants No. 11904008 and No. 11847002); the Natural Science Foundation of Anhui Province (Grant No. 1908085QA21); the China Postdoctoral Science Foundation (Grant No. 2019M662132). H. W. W. thanks Hai-Long He of Wuhan University for helpful discussions.

## Author contributions

H. W. W. conceived the idea and did the theoretical analyses. J. Q. Q. performed the numerical

simulations. J. Q. Q and Y. K. L. performed experimental measurements. H. W. W., Y. P., Z. Q. S. and L. W. J. guided the research. All authors contributed to discussions of the results and paper preparation. H. W. W. and J. Q. Q wrote the paper.

**Competing interests**

The authors declare no competing interests.

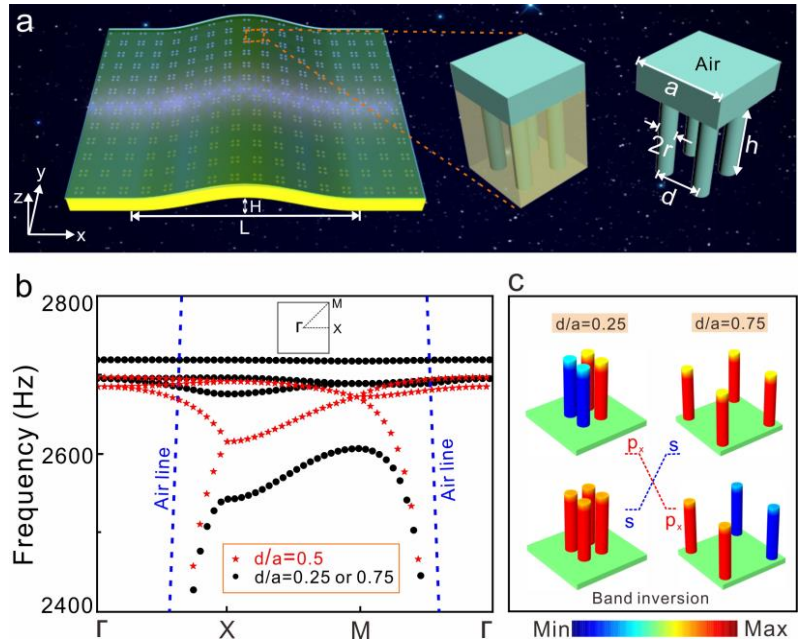

**Fig. 1| Unit cell of 2D curved acoustic TI and band inversion**. **a** Schematic of curved second-order TI by perforation on curved rigid plate with height *H* and span *L*. The right insets show the details of the unit cell as the lattice length *a*, hole radius *r*, hole depth *h*, and hole distance *d*. The yellow and cyan areas represent the rigid wall and air, respectively. The interface between the topological trivial and nontrivial lattices is indicated by the light trace. **b** Band structure of square lattice for different ratios *d*/*a* = 0.5 (red pentagrams) and 0.25/0.75 (black dots). The blue dashed lines represent air lines and the inset shows the first Brillouin zone. **c** Band inversion between topological trivial phase (*d*/*a* = 0.25) and nontrivial phase (*d*/*a* = 0.75). The colors correspond to the pressure distributions at point X.

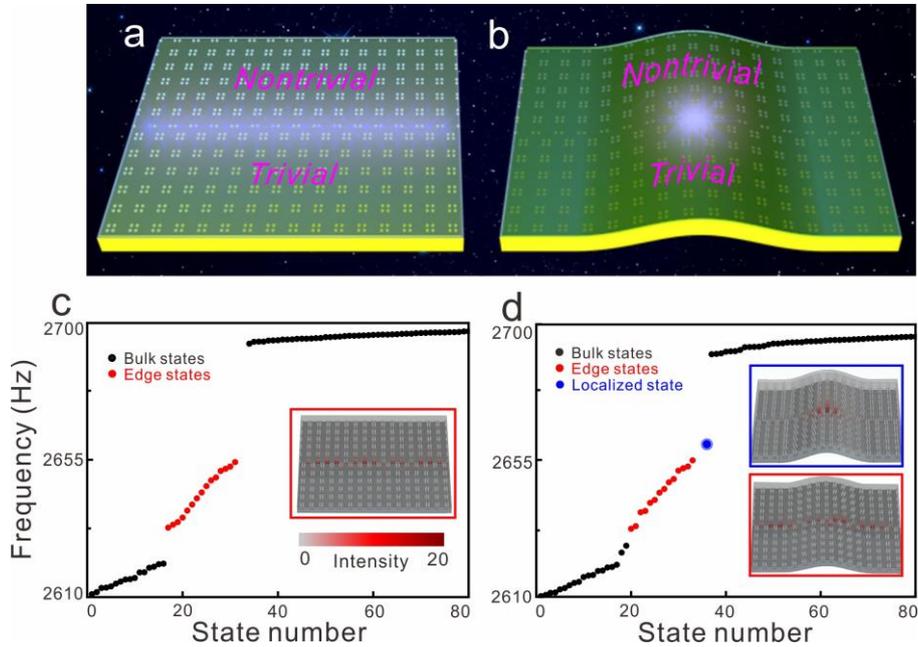

**Fig. 2| Topological localized state induced by geometrical curvature.** Structural diagrams for **a** flat and **b** curved holey TIs. The eigenfrequency distributions for the flat and curved structures are depicted in **c** and **d**, respectively. The blue, red, and black dots represent the localized, edge, and bulk states, respectively. The insets with the red boxes in **c** and **d** correspond to the pressure distributions of the 12th edge states of the flat and curved structures, respectively. The inset with the blue box in **d** represent the pressure distribution of the topological localized state.

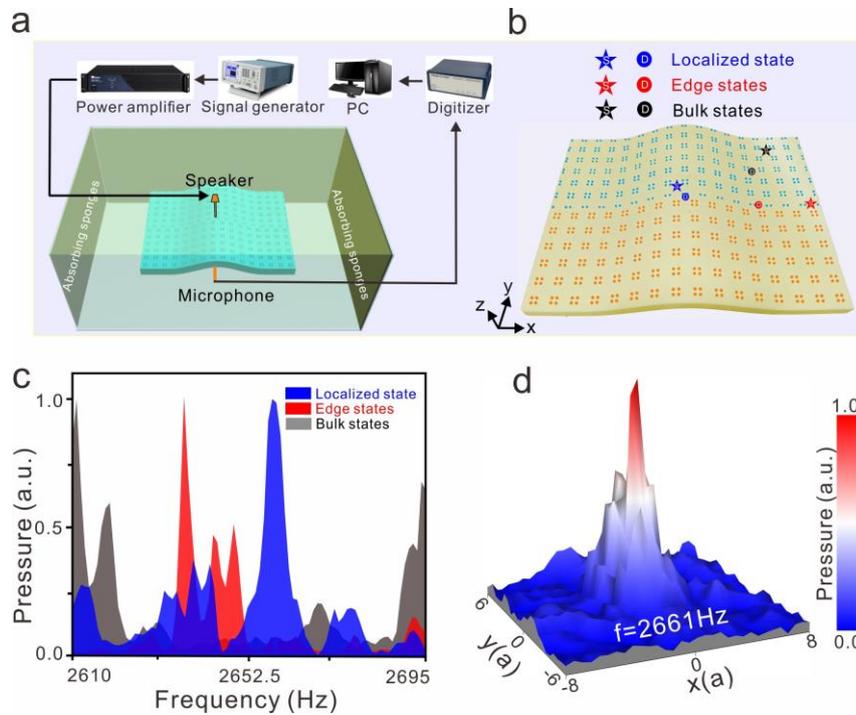

**Fig. 3| Experimental observation of topological localized state. a** Schematic of experimental setup. **b** Photograph of sample, including topological trivial part (orange holes) and nontrivial part (cyan holes). The blue, red, and black stars represent the positions of the point source and the corresponding dots indicate the positions of the microphone for measuring the pressure spectra. **c** Measured pressure spectra for bulk, edge, and localized states. **d** Measured pressure distribution of localized state with frequency $f$ = 2661 Hz in all holes of convex surface.

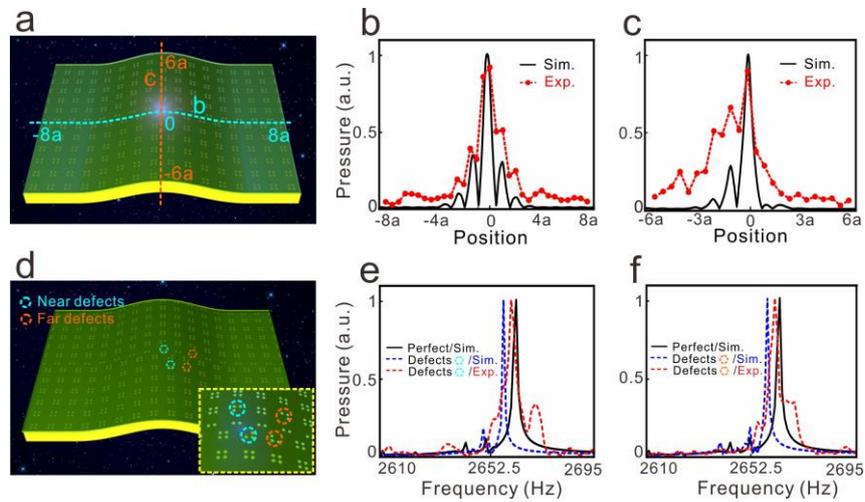

**Fig. 4| Robustness of topological localized state against defects.** Measurements of the pressure distributions along the cyan and orange dashed lines b and c in **a**. The simulated (black solid line) and experimentally measured (red dot-dashed line) results along b and c are shown in **b** and **c**, respectively. **d** Schematic for introducing near and far defects into curved TI, indicated by cyan and orange circles. The inset shows the amplification of the structural top. Transmission spectra for the structure including **e** near defects and **f** far defects obtained by simulation (blue dashed line) and experiment (red dashed lines). The black solid lines represent the results of the perfect structure as a reference.

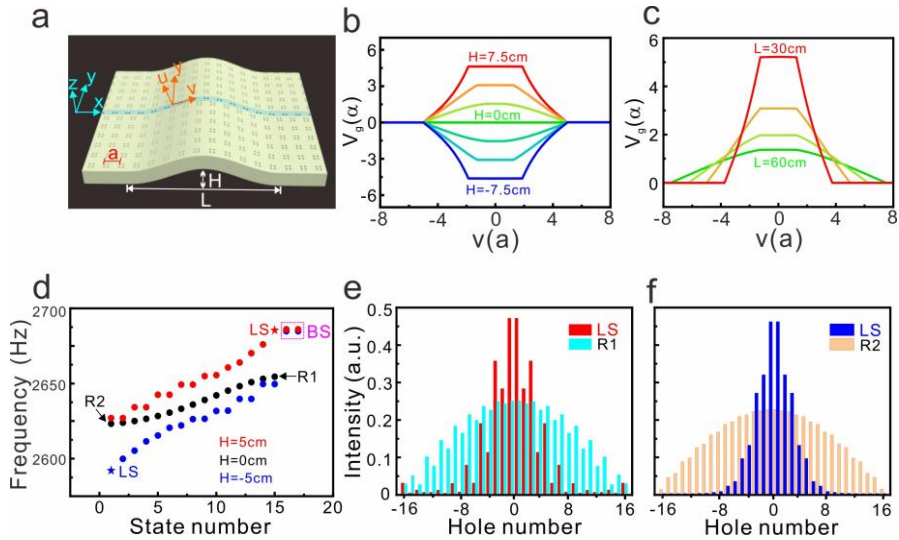

**Fig. 5| Theoretically calculated geometric potentials and eigenstate distributions of curved structure. a** Geometry and coordinate system of 2D curved structure, where ($x$, $y$, $z$) and ($v$, $y$, $u$) represent the Cartesian and curvilinear coordinates, respectively, $H$ and $L$ correspond to the height and span of the bump, respectively, and $a$ is the lattice constant. Geometric potential as function of curvilinear coordinate $v$ for different **b** heights and **c** spans. **d** Eigenfrequencies of 1D hole array (as marked by cyan strip in **a**) for different heights $H$ = 5 cm (red dots), 0 cm (black dots), and -5 cm (blue dots) based on tight-binding model. The top right dashed box depicts the boundary states, and the red and blue pentagrams marked with "LS" indicate the topological localized states for heights $H$ = 5 cm and -5 cm, respectively. "R1" and "R2" indicate the eigenstates with the maximal and minimal frequencies of the flat structure as references. Comparison of mode distributions of localized state marked by red "LS" and reference "R1" in **e**, and localized state marked by blue "LS" and reference "R2" in **f**.

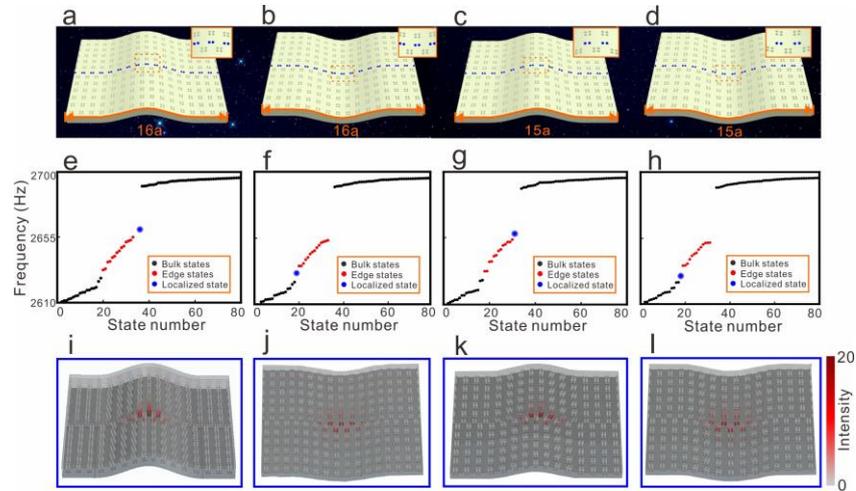

**Fig. 6| Eigenfrequencies and mode distributions for different curved directions and lattice distributions. a** to **d** Schematics of 2D curved structures, where **a** and **b** depict 16 unit cells along the curved surface but different curved directions, **c** and **d** show 15 unit cells along the curved surface with different curved directions, and the top-right orange dashed boxes indicate the details of the lattice distribution at the top or bottom of the curved surface. **e** to **h** present the eigenfrequencies of the curved structures in **a** to **d**, respectively, and **i** to **l** depict the mode distributions of the topological localized states supported in the corresponding curved structures.

*Supplementary Materials*

# Observation of protected localized states induced by curved space in acoustic topological insulators

Hong-Wei Wu,[1,2,*] Jia-Qi Quan,[1] Yun-Kai Liu,[1] Yiming Pan,[3] Zong-Qiang Sheng[1] & Lai-Wang Jing[1]

[1]*School of Mechanics and Photoelectric Physics, Anhui University of Science and Technology, Huainan 232001, China*
[2]*National Laboratory of Solid State Microstructures, Nanjing University, Nanjing 210093, China*
[3]*Physics Department and Solid State Institute, Technion, Haifa 32000, Israel*
∗Correspondence and requests for materials should be addressed to <hwwu@aust.edu.cn> (HWW)

1. **Geometric potential effect in 2D curved TI structure.**

In Fig. 5a of main text, we perform a conformal transformation $z + ix = R(v) \exp[(u + iv)/R(v)]$ between cartesian coordinates $(x, y, z)$ and curvilinear coordinates $(v, y, u)$, where $R(v)$ represents the radius of curvature varied with curvilinear coordinate $v$. Based on the conformal transformation, the 3D scalar wave equation for the envelope function of the pressure field $p(v, y, u)$ in curvilinear coordinates can be expressed as[1,2]:

$$\frac{\partial^2 p}{\partial v^2} + \frac{\partial^2 p}{\partial u^2} + \exp\left(\frac{2u}{R(v)}\right)\left(\frac{\partial^2 p}{\partial y^2} + k^2(v, y, u)p\right) = 0, \quad \text{(S1)}$$

where $k(v, y, u)$ is the wavevector for the sound wave propagation in the holey material. For weak curvature such that $R(v) \gg u$, one can assume $\exp(2u/R(v)) \approx 1 + 2u/R(v)$. In our curved second-order TI structure, it is well known that the waves only propagate along the interface between the topological trivial and nontrivial parts due to the confinement of the bulk bandgap. Thus, the Eq. (S1) can be simplified to a 2D scalar wave equation as:

$$\frac{\partial^2 p}{\partial v^2} + \frac{\partial^2 p}{\partial u^2} + k^2(v,u)p + \frac{2u}{R(v)}k^2(v,u)p = 0. \quad \text{(S2)}$$

For the spoof acoustic surface waves supported by the one-dimensional holey structures as marked by cyan curved strip in Fig. 5a, the surface wave can be written as $p(v, u) = A(v, u)\exp(i\beta v)$ under the paraxial approximation, where $A(v, u)$ is the slowly varying amplitude and $\beta$ is the wavevector. By applying the slowly varying envelope approximation on 2D scalar wave equation, the acoustic surface wave yields the

acoustic Schrödinger equation[3,4]:

$$i\hbar \frac{\partial A}{\partial v} = -\frac{\hbar^2}{2n_{sw}} \frac{\partial^2 A}{\partial u^2} + \frac{n_{sw}^2 - n^2(v,u)}{2n_{sw}} A - \frac{un^2(v,u)}{R(v)n_{sw}} A, \quad (S3)$$

where $\hbar = 1/k_0$, $k_0$ is the wavevector of free space. Here, $n(v, u) = k(v, u) / k_0$ corresponds to the refractive index of holey materials, and $n_{sw} = \beta / k_0$. For spoof acoustic surface waves confined on the curved surface, the second term in right side of Eq. (S3) corresponds to confining potential $V_c(v, u) = [n_{sw}^2 - n^2(v, u)] / (2n_{sw})$. The last term is the geometric potential $V_g(v, u) = - un^2(v, u) / R(v)n_{sw}$. Here, the geometric potential is mainly depended on the curvature $R(v)$ and can be expressed as $V_g(v) = - \alpha / R(v)$ for the spoof surface wave propagated on the curved surface, $\alpha = un^2(v, u) / n_{sw}$ as a parameter.

The Hamiltonian of 1D SSH model system for a finite coupled hole array can be written as[5]:

$$H = \begin{pmatrix} \omega_0 + V_{g1} & c & & & & \\ c & \omega_0 + V_{g2} & t & & & \\ & t & \ddots & t & & \\ & & t & \omega_0 + V_{g31} & c \\ & & & c & \omega_0 + V_{g32} \end{pmatrix}, \quad (S4)$$

where the resonant frequency of the hole is $\omega_0 = 2685$Hz, and the coupling coefficients between two holes are $t = 50$Hz for the ratio $d/a = 0.25$ and $c = 15$Hz for the ratio $d/a = 0.75$ obtained by simulation. The discrete geometric potentials $V_g$ are added to the resonant frequencies of holes distributed at different positions of curved surface. In our structure, the geometric potentials have the relationship: $V_{g1} = V_{g32}$, $V_{g2} = V_{g31}$, …, $V_{g16} = V_{g17}$ due to the symmetry of structure. Then we can calculate the eigenvalues and eigenvectors of the Hamiltonian simply.

## 2. Eigenfrequency distributions of 2D flat topological insulators for different

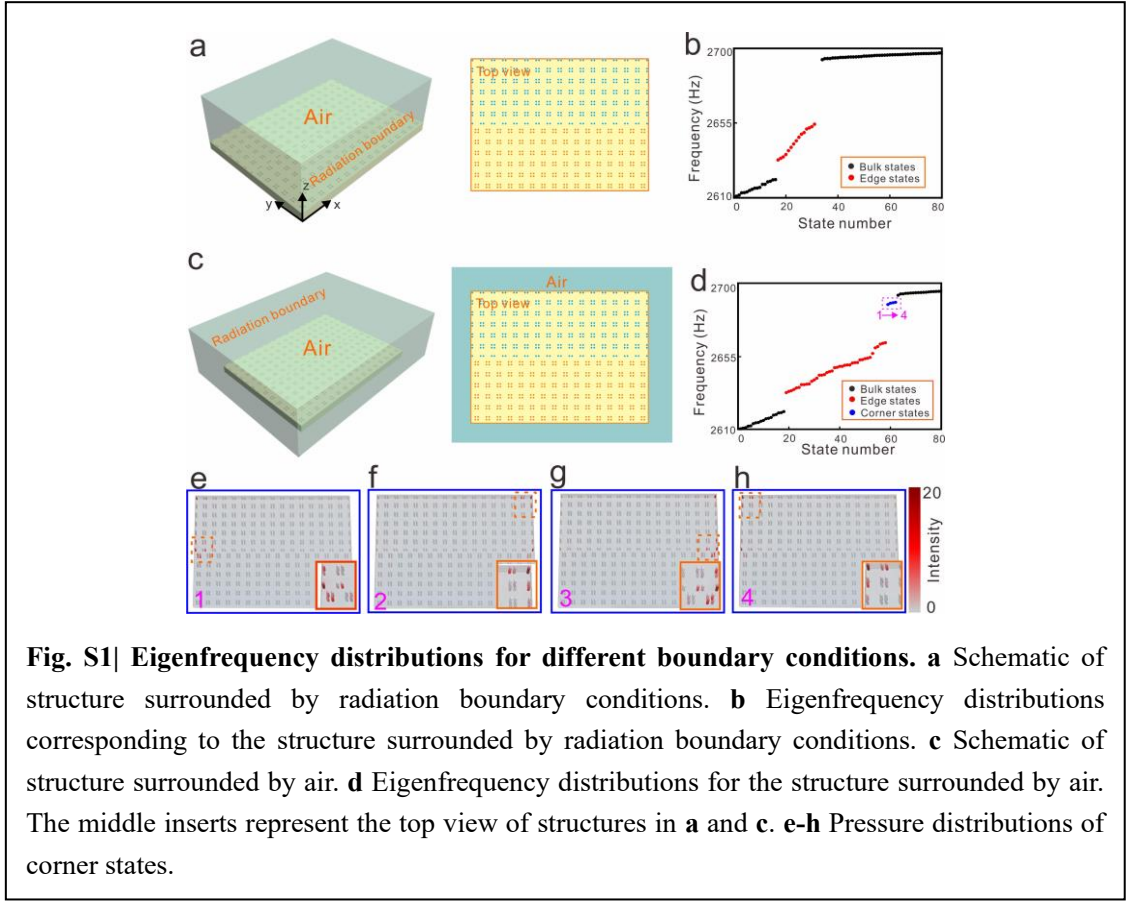

**Fig. S1| Eigenfrequency distributions for different boundary conditions. a** Schematic of structure surrounded by radiation boundary conditions. **b** Eigenfrequency distributions corresponding to the structure surrounded by radiation boundary conditions. **c** Schematic of structure surrounded by air. **d** Eigenfrequency distributions for the structure surrounded by air. The middle inserts represent the top view of structures in **a** and **c**. **e-h** Pressure distributions of corner states.

environments .

It is well known that the open boundary condition is required for the existence of boundary states in topological insulators. For finite-size topological insulators, different surrounded materials induce different eigenfrequency distributions. Generally, the topological trivial lattices with same gap width are arranged around the topological nontrivial lattices for clearly observing the topological edge and corner states. In our paper, we want to investigate the influence of curved surface on the edge state distributed along the interface between the topological nontrivial and trivial parts. Thus, we set the boundary conditions at the outside boundaries as plane wave radiation boundary conditions. In that case, the edge and corner states supported at the outside boundaries of topological nontrivial lattices will be suppressed, and only present the edge states along the interface. To clarify this point, we calculate the eigenstate distributions for the topological flat structure around by radiation boundary conditions and air as can be seen in a and c of Fig. S1. In Fig. S1, b and d show the eigenfrequency distributions. We can find that the structure for applying radiation boundary conditions at outside boundaries only supports the edge states along the middle interface in the bulk gap in Fig. S1b. However, the structure surrounded by air not only supports the edge states along the middle interface, but also supports the edge states along the other three boundaries and the corner states at the four corners in the nontrivial part. Thus, we can find that more edge states and four corner states present in the bulk gap in Fig. S1d. Figs. S1e-h show the pressure distributions of corner states corresponding to the blue dots in d. Right-bottom inserts show the amplifying mode distribution marked by dashed box.

## 3. Pressure distributions of the edge states for flat and curved structure in simulations.

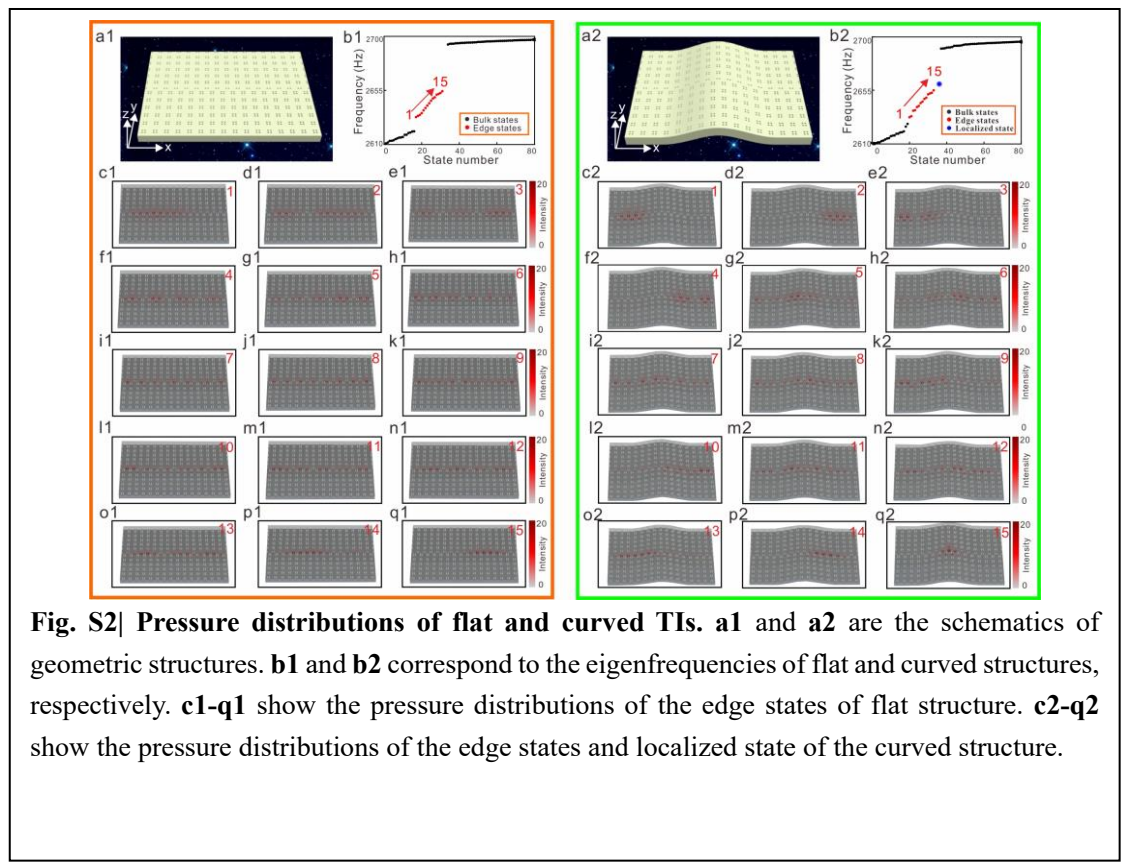

**Fig. S2| Pressure distributions of flat and curved TIs. a1** and **a2** are the schematics of geometric structures. **b1** and **b2** correspond to the eigenfrequencies of flat and curved structures, respectively. **c1-q1** show the pressure distributions of the edge states of flat structure. **c2-q2** show the pressure distributions of the edge states and localized state of the curved structure.

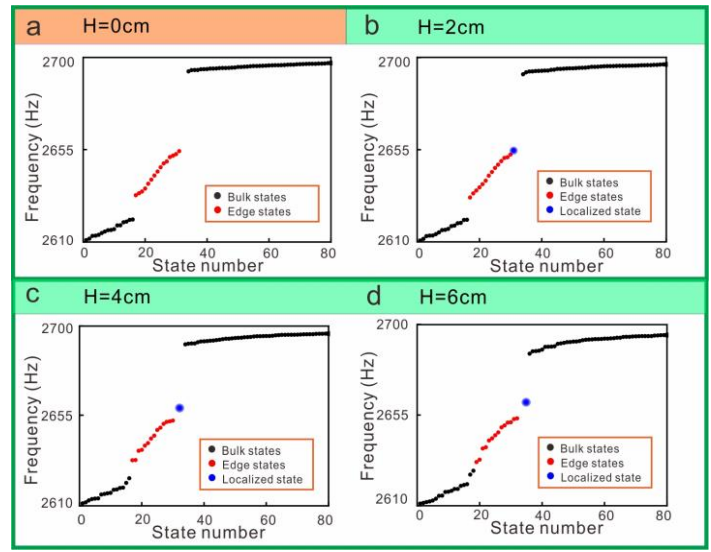

**Fig. S3| Evolution of eigenfrequencies with increasing the curved height $H$. a** Eigenfrequency distributions for the flat structure, red dots and black dots represent edge and bulk states. **b-d** Calculated eigenfrequencies for increasing curved height from $H$ = 2cm to 6cm. A topological localized state marked by blue dot stands out from the edge states.

4. Evolution of the eigenfrequency with increasing the curved height H.

5. Mode distributions of eigenstates of 1D hole array for different geometric curvatures.

By considering the geometric potentials, the eigenfrequencies and eigenstates of 1D hole array distributed on the curved surface can be obtained by adding the discrete geometric potentials to the onsite energy of holes. Then, the eigenfrequency distributions are shown in Fig. 5d of main text for structures with different curved height H = 5cm, 0cm, -5cm. Here, we show the eigenstate distributions corresponding to the eigenfrequencies. The results in Figs. S4, S5 and S6 correspond to the black, red and blue dots in Fig. 5d.

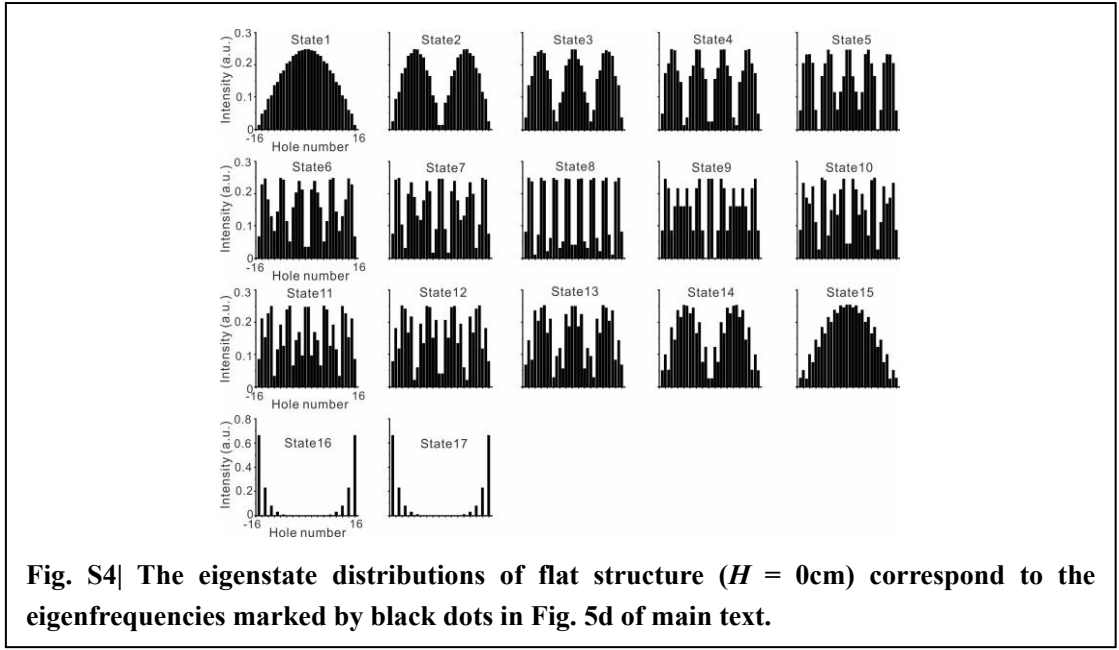

**Fig. S4| The eigenstate distributions of flat structure ($H = 0$cm) correspond to the eigenfrequencies marked by black dots in Fig. 5d of main text.**

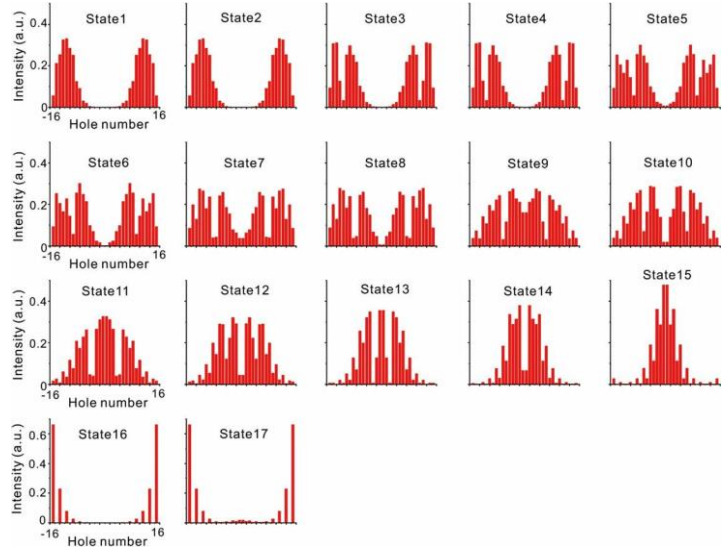

**Fig. S5|** The eigenstate distributions of convex curved structure ($H$ = 5cm) correspond to the eigenfrequencies marked by red dots in Fig. 5d of main text.

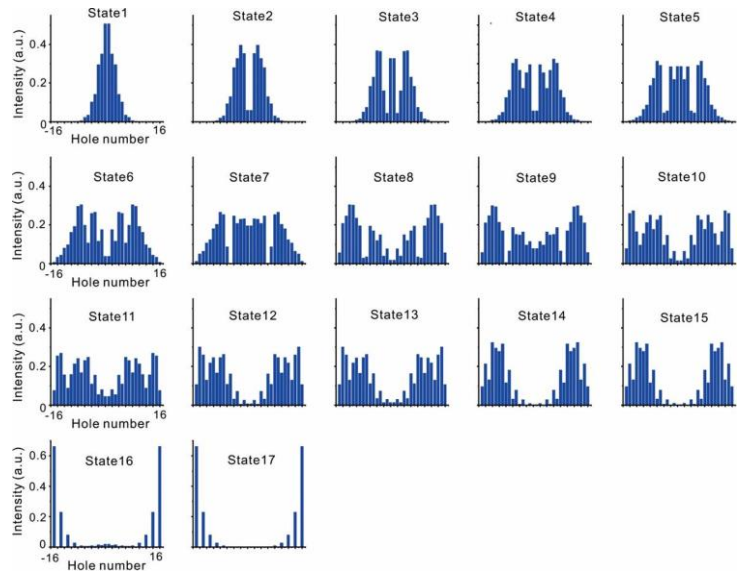

**Fig. S6|** The eigenstate distributions of concave curved structure ($H$ = -5cm) correspond to the eigenfrequencies marked by blue dots in Fig. 5d of main text.